\documentstyle[prl,aps,epsfig,multicol]{revtex}

\newcommand{\be}{\begin{eqnarray}}
\newcommand{\ee}{\end{eqnarray}}

\begin{document}
\draft

\title{Quantum phase retrieval of a Rydberg wave packet using a half-cycle pulse}

\author{J.~Ahn, D.~N.~Hutchinson, C.~Rangan, and P.~H.~Bucksbaum}
\address{Physics Department, University of Michigan, Ann Arbor, MI 48109-1120}

\maketitle

\begin{abstract}
A terahertz half-cycle pulse was used to retrieve information
stored as quantum phase in an $N$-state Rydberg atom data
register. The register was prepared as a wave packet  with one
state phase-reversed from the others (the ``marked bit'').  A
half-cycle pulse then drove a significant portion of the electron
probability into the flipped state via multimode interference.
\end{abstract}


\begin{multicols}{2}
\narrowtext

An atomic Rydberg wave packet is an atom with one of its electrons
in a coherent superposition of many high-lying orbitals. Wave
packets can be viewed as data registers for the information
contained in the relative quantum phase of their constituent
orbitals. One problem is that phase is difficult to
detect. State selective detection on an ensemble of identically
prepared registers usually only reveals the component amplitudes;
the phases are hidden. Optical techniques have been employed to
store and then efficiently retrieve phase information from atomic
Rydberg wave packets~\cite{AhnScience00}, following a suggestion
that binary information stored as phase in quantum registers
could be searched more efficiently than the same classical
information in classical data
registers~\cite{GroverPhysRevLett97ab}. This Letter reports a
different technique for extracting phase information, based on
the interaction between the wave packet and a broadband coherent
electric field pulse.

In the experiment reported in Ref.~\cite{AhnScience00}, a shaped optical
pulse carried binary information into the Rydberg atom, by
creating an electron wave packet with one or more states
phase-flipped (binary 1) with respect to the others (binary 0). A
second optical pulse created another 
wave packet (the decoder) 
which holographically interfered with the register wave 
packet~\cite{Wavepacketholography,Quantumsculpting}.
This interference converted the phase
information to amplitudes by amplifying the flipped states and
suppressing the unflipped states. 
This method of information
storage is limited because it depends on mutual coherence between
a low-lying launch state and the Rydberg states. \ Furthermore,
the decoding pulse requires that most of the electron's
probability amplitude resides in the launch state, and only a
small portion goes into Rydberg states. \ Therefore, the
scheme demonstrated in the above experiment cannot be scaled to a
very large data register~\cite {MeyerScience00}.

In the new data storage and retrieval experiment presented in
this paper, the launch state is no longer part of the data
register, and all of the probability resides in the Rydberg
states of the atom. \ The decoding process of amplifying the
flipped bits is now performed by a terahertz domain half-cycle
pulse(HCP)~\cite{HCPgeneration}. This broadband coherent
far-infrared radiation directly couples together Rydberg levels.
Information stored in the phase domain is converted to amplitude
information by coherent redistribution induced by the HCP.

HCP interactions with the Rydberg atoms have been studied
previously, in HCP-ionization experiments and model
calculations~\cite {HCPionization,Impulsivemodel}, HCP-redistribution of energy
eigenstates~\cite{TielkingPRA95}, and interactions with Rydberg
wave packets~\cite{RamanPRL96,WavepacketHCPinteraction}. In the
present work, an HCP amplifies a phase-flipped orbital (the
marked bit), suppressing the other unmarked states in the wave
packet.

The HCP locates the marked bit because its activity depends on the
wave packet's location. \ The interaction is strongest when the
wave packet is localized at the ion core, and this situation
corresponds to a coherent sum of orbitals with equal
phase~\cite{RamanPRL96}.  A Rydberg wave packet data register
with one flipped bit is described by a superposition of such a
localized wave packet, plus a phase-reversed orbital:
\begin{eqnarray}
|\Psi _{i}>=\sum_{j}|n_jlm\rangle -2|n_{o}lm\rangle
\label{flippedbit}
\end{eqnarray}
A relatively weak HCP interaction will destroy the localized
component easily, while, the spatially delocalized single marked
orbital $|n_{o}lm\rangle $ is amplified by a factor of
approximately $2^{2}=4.$ This is an example of the quantum
diffusion and quantum amplification of information, discussed by
Grover~\cite{GroverScience98}.

The database search experiment was performed in a beam of atomic
Cs. The Rydberg data register was loaded using wave packet
sculpting techniques that have been described previously~\cite
{Quantumsculpting}. \ Briefly, an optical pulse was sculpted with
an acousto-optic Fourier filter~\cite{Pulseshaper} to excite a
Rydberg data register with one marked bit in Cs $p_{z}$-states
with n=24-29. State-selective field ionization~(SSFI)~\cite {SSFI}
of the initial wave packet reveals only the amplitudes of the
states, as shown in Fig.~1(a).

An HCP was generated inside the atomic beam apparatus by an
ultrafast optical pulse illuminating a large aperture ($1.5\times
3cm^{2}$) electrically biased GaAs wafer \cite{HCPgeneration}. The
far-infrared radiation was focused on the atomic beam by an
off-axis parabolic mirror. Its peak field strength was calibrated
by observing the threshold for ionization of Rydberg states
\cite{HCPionization}. The HCP delivered an impulse of
approximately $Q_{o}=0.0043 \hbar /a_{0}$, which was enough to
significantly mix $s$, $p$, and $d$ states \cite{TielkingPRA95}.
It was polarized parallel to the Rydberg state.

The HCP interrogated the system at time $\tau $, converting the
wave packet to a simpler structure in which only the marked state
remains.  Figure 1 shows SSFI spectra for three different data
register configurations, in which a different state in the
register was phase-reversed as shown in Eq.~(\ref{flippedbit}):
the flipped bit was $25p,$ $26p$, and $27p$ in panels (ii), (iii),
and (iv), respectively.

The quantum diffusion and marked state amplification can be calculated in the limit
where the duration of the HCP is much less than the natural time scale for the phase
evolution of the Rydberg wave packet, i.e. the Kepler orbit time, $t_{%
{\rm K}}=2\pi /\Delta E=2\pi (n-\delta _{l})^{3}$, where $n$ is the
principal quantum number and $\delta _{l}$ is the quantum defect for the $%
l^{th}$ angular momentum states. For atomic cesium at $n=26$,
$t_{{\rm K}}\approx $ 2ps is five times the full width half
maximum (FWHM) of the HCP $\tau _{{\rm FWHM}}\approx$ 440fs. In
this impulsive limit, the HCP atomic interaction Hamiltonian (in
atomic units with $e=m_e=\hbar=1)$
\begin{eqnarray}
{\mathcal H}(t) =& \vec{E}(t)\cdot \vec{r} & = - \vec{Q}\delta
(t)\cdot \vec{r},
\end{eqnarray}
where $\vec{Q}$ $=\int_{t-\epsilon }^{t+\epsilon
}\vec{E}(t^{\prime })dt^{\prime }$ is the total momentum transfer
to the electron. The initial electron wave function $|\Psi
_{i}\rangle $ acquires an extra phase along the direction of the
impulse($\hat{z}$) 
\be
|\Psi _{f}\rangle&=&e^{i\overrightarrow{Q}\cdot
\overrightarrow{r}}|\Psi _{i}\rangle . 
\ee
For an initial wave packet
$|\Psi _{i}\rangle=\sum_{n}a_{n}|n l m\rangle$ , the final state
wave packet is 
\be
|\Psi _{f}\rangle
&=&\sum_{n'l'm',n}a_{n}f_{nlm}^{n'l'm'}|nlm\rangle, 
\ee
\noindent where the
transition matrix elements are given by 
\be
f_{nlm}^{n'l'm'}
&=&\langle n'l'm'|e^{i\overrightarrow{Q}\cdot \overrightarrow{r}%
}|nlm\rangle .  
\ee
\noindent For $l=1$ and $m=0$, and $%
\overrightarrow{Q}$ directed along $\widehat{z}$, we find
\begin{eqnarray}
f_{np0}^{n'l'm'}(Q)&=&\langle R_{n'l'}(r)| i^{l'-1}
\sqrt{\frac{3}{2l'+1}}\left[ l'j_{l'-1}(Qr)\right. \nonumber \\
& &\left. -(l'-1)j_{l'+1}(Qr)\right] |R_{np}(r)\rangle \delta
_{m',0}.
\end{eqnarray}
The phase of each transition $f_{nlm}^{n'l'm'}(Q)$ is determined
by the parity change, $\triangle l=l'-l$ and the magnitude of $Q$.
For example, $p $-$p$ transitions are always real and $p$-$d$
transitions are imaginary, while the sign changes with $Q$.

The probability density in a final state $\left|
n'l'm'\right\rangle $ is given by
\begin{eqnarray}
P_{n'l'm'}=\left| \sum_{n}a_{n}f_{nlm}^{n'l'm'}(Q)\right|^{2}.
\end{eqnarray}
For atomic cesium levels around $n\approx 26$ excited by an
impulse $Q=Q_{0}$, the self-transition amplitude $f^{np0}_{np0}$
is $\pi$ out of phase with transition amplitudes from other
orbitals $f^{n'p0}_{np0}$. This creates constructive
multi-mode interference for the marked states, while transitions
to other states experience destructive interference. The results
of a calculation in the impulse limit 
are shown in Fig.~1, and show qualitative agreement.
Deviations from the impulse limit were studied with a second
calculation, where the Schr\"{o}dinger equation was integrated
over a finite-duration model half-cycle pulse. We used an HCP
model fitted to experimentally measured
spectra~\cite{Impulsivemodel}.  The momentum transferred by
the HCP, $Q$, is the area under the curve,
\begin{eqnarray}
E(t) &=&\left\{ \begin{array}{ll}
0,&t<0 \\
29.56E_{\rm peak}\left[ 17.75\left(t/\tau\right)^{3}e^{-8.87
t/\tau} \right.
& \\
\left. -0.412 \left(t/\tau\right)^{5}e^{-4.73 t/\tau }\right] ,&
t\geq 0
\end{array} \right\}.
\end{eqnarray}

The field-free basis states were calculated using a grid-based
pseudopotential method \cite{Pseudopotential}. The time-evolution 
was carried out by using 
a split-operator method in a restricted
basis of essential states \cite{Splitoperator}. This symmetrized-product propagator is
unitary and correct through the second-order in the time step. We
used a
basis of 187 states, that is, states with quantum numbers $n=21-31$ and $%
\ell <17(m=0)$ and a time step of 10fs. The results are shown in Fig.~1. When a
single initial state out of the wave packet is phase-flipped, we reproduce the result
that this orbital survives the HCP with a significantly larger population than the
rest, in good agreement with the experiment.

The full dynamics of the data base are revealed by mapping the HCP
redistribution as a function of the target time. Since the
Rydberg levels in the data register are non-degenerate, different
configurations of flipped bits occur at times other than $\tau$.
\ This two-dimensional phase space interference pattern, or
``quantum carpet''\cite{Quantumcarpet}, is shown in Fig.~2, for
data collected over several Kepler periods. \ The HCP consistently
redistributed the wave packet into a single state when it
consisted of only a single flipped orbital. The patterns of
ridges in the quantum carpet show how the Kepler period changes
with energy, and also display wave packet dispersion over 8 ps of
the evolution.  The lines are equivalent to classical ridges of a
multi-mode interference pattern\cite{Quantumcarpet}.

A quantitative measure of success in the database search problem
is the reduction of the informational entropy of the system
following the search algorithm \cite{Shannontheory}. Entropy $S$
in this simple case of a single marked state is defined as
\be\label{entropy} S=-\sum_{i=1}^{N}P_{i}\log P_{i}, \ee where
$P_{i}$ is the probability that the marked bit resides in orbital
$i$. The information system consists of the $N$-state data
register. When the data register is loaded, the information is
encoded as the phase of each orbital. The orbital amplitudes
carry no information. Since SSFI only measures amplitudes, the
information is completely hidden from view, and therefore the
entropy of the system is maximum: $S_{\rm initial}=-\log (1/N) = \log N$.

The interaction with the HCP converts the phase information into
amplitudes, which can be read out using SSFI. The rules of
quantum measurement insure that only one state will be detected
each time an atom is interrogated by SSFI. If the algorithm works
perfectly, the marked orbital is revealed with unit efficiency,
and the entropy drops to zero.  Otherwise, unmarked orbitals may
occasionally be detected by mistake. The entropy reduction can be
calculated from the probability distribution, using
Eq.~(\ref{entropy}).

The states that are not included in the data register can rob
some of the probability during the HCP operation, and therefore
contribute to the inefficiency of the algorithm. A lower bound on
the amount of entropy contributed by leakage into the
non-register states can be measured by including a reservoir
state in the calculation of the entropy. Table 1 shows the
entropy reduction for each of the marked orbitals shown in Fig.~1.
 There is a significant entropy reduction, but the algorithm is
far from perfect. Further progress could be made by shaping
broadband terahertz radiation (which has been demonstrated using
optical pulse shaping techniques \cite{ShapingThz}) to focus the
state redistribution.

In conclusion, we have used a HCP to search for information
stored as phase in a Rydberg wave packet.  Phase information
coded in a Rydberg quantum register was retrieved through the
impulsive interaction of HCP. This interaction forced a
constructive multi-mode interference only on a phase-flipped
state. Direct integration of the time-dependent Schr\"{o}dinger
equation as well as an impulsive model calculation showed good
agreements with the experimental data. Beyond
the problem of database searching, these results point toward the
use of HCPs as general tools of analysis for unknown wave
packets. \ This could be used in conjunction with optical
techniques for wave packet 
tomography~\cite{Quantumsculpting}.

It is a pleasure to acknowledge useful discussions and help from
T.~C.~Weinacht, J.~L.~White P.~R.~Berman, K.~J.~Schafer, and
R.~R.~Jones. This research was supported by the NSF grant
9987916, and the Army Research Office. \ One of us (C.R.)
gratefully acknowledges support from the Fellows and Visitors
program of the Center for Ultrafast Optical Science at the
University of Michigan.


\begin{table}[tbp]
\label{table1}
\caption{Informational entropy calculations using Eq.~\ref{entropy} for 
experimental data $S_{\rm Exp}$, the impulse 
model calculation $S_{\rm Imp}$, and the 
full calculation with the time-integration of the Schr\"{o}dinger 
equation $S_{\rm Full}$.
The initial entropy, $S_{\rm initial}=1.79$ for $N=6$.}
\begin{tabular}{cccc}
  Marked bit & $S_{\rm Exp}$
& $S_{\rm Imp}
$ & $S_{\rm Full}$
\\
\tableline
  25p & 1.09 & 1.004 & 1.116 \\
  26p & 0.56 & 1.032 & 1.081 \\
  27p & 1.17 & 1.071 & 1.037 \\
\end{tabular}
\end{table}

\begin{figure}[tbp]
\label{fig1}
\centering
\caption{Experimental data and model calculations of the 
state-selective field ionization spectra of the initial wave
packet(i), and the final wave packets at $\protect\tau=2.1$ps(ii),
$4.2$ps(iii) and $4.7$ps(iv). (a) Experimental data (b) Impulse
model calculation (c)
Model calculation includes the direct
integration of the time dependent Schr\"odinger equation using
the split-operator method.  
Both calculations assumed an initial
Gaussian wave packet $|\psi_{t=0} \rangle= 0.5|24p\rangle+
|25p\rangle + 1.2|26p\rangle +|27p\rangle +0.7|28p\rangle
+0.5|29p\rangle$. The electron probability of the initial
Gaussian wave packet was driven into the 25p(ii), 26p(iii), and
27p(iv) states respectively. The nearly degenerate $p$ and $d$ states and 
the $s$ and hydrogen-like states are not resolved in the experimental data.}
\rotatebox{0}{\includegraphics[scale=0.55]{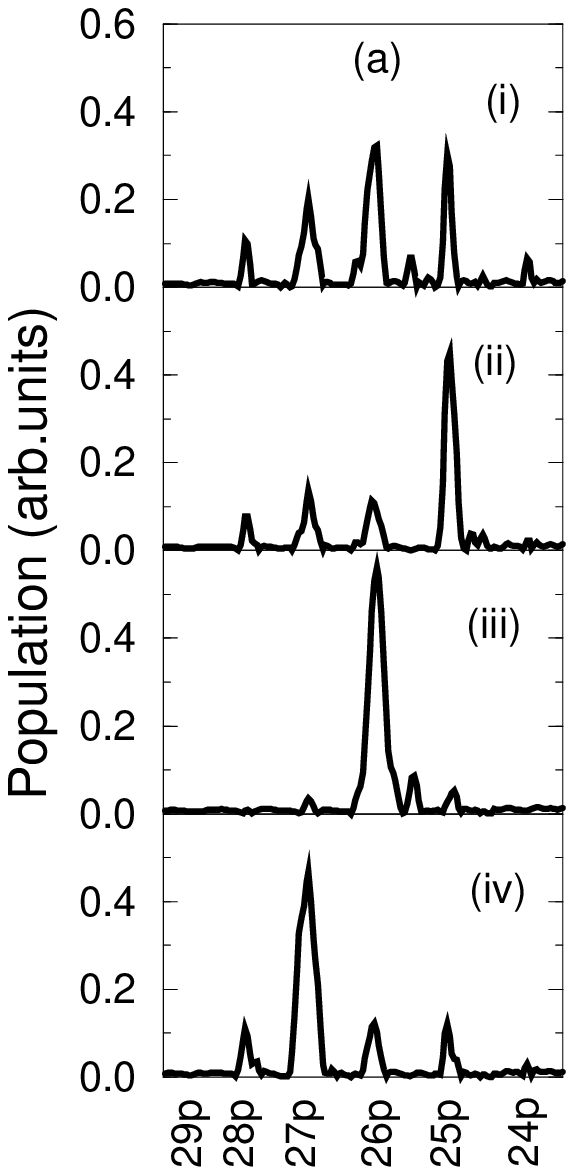}}
\rotatebox{0}{\includegraphics[scale=0.55]{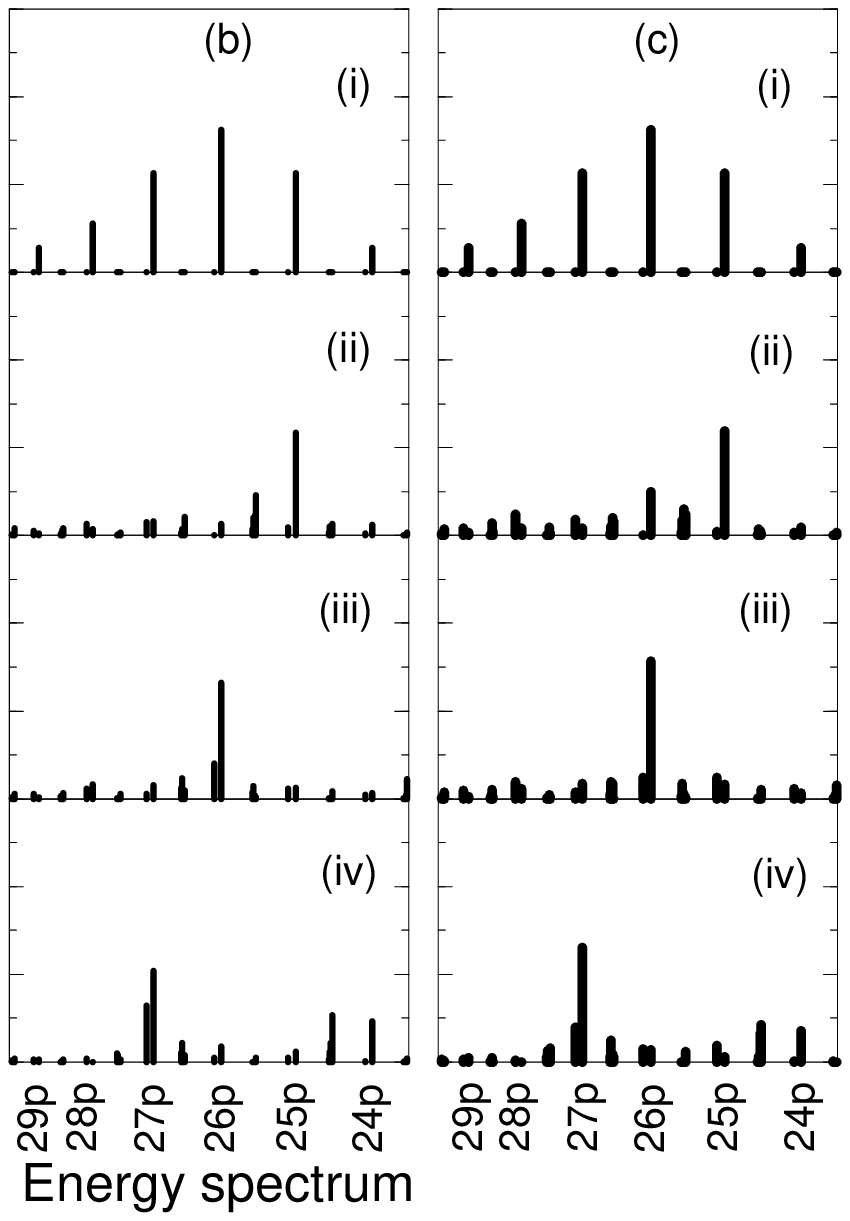}}
\end{figure}


\begin{figure}[tbp]
\label{fig2}
\centering
\caption{Experiment(a) and theory(b) of the multi-mode
interference pattern of Rydberg wave packets forced by the
half-cycle pulse interaction. Guide lines follow the phase
gradient maxima that appear around the half multiples
of the Kepler orbit times $\protect\tau=(k+1/2)\protect t_{\rm K}(n)$ with $%
k=0,1 $ and $2$. }
\includegraphics[scale=0.55]{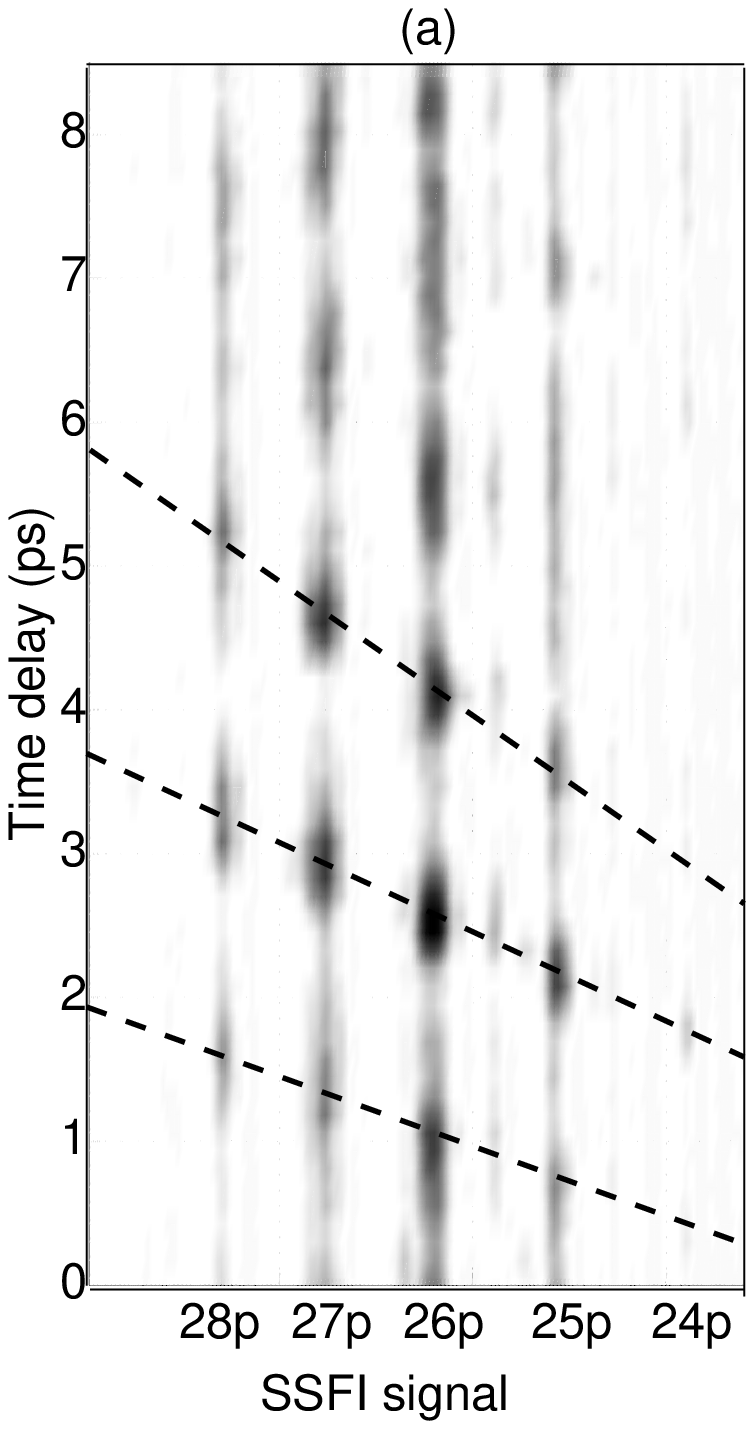}
\includegraphics[scale=0.55]{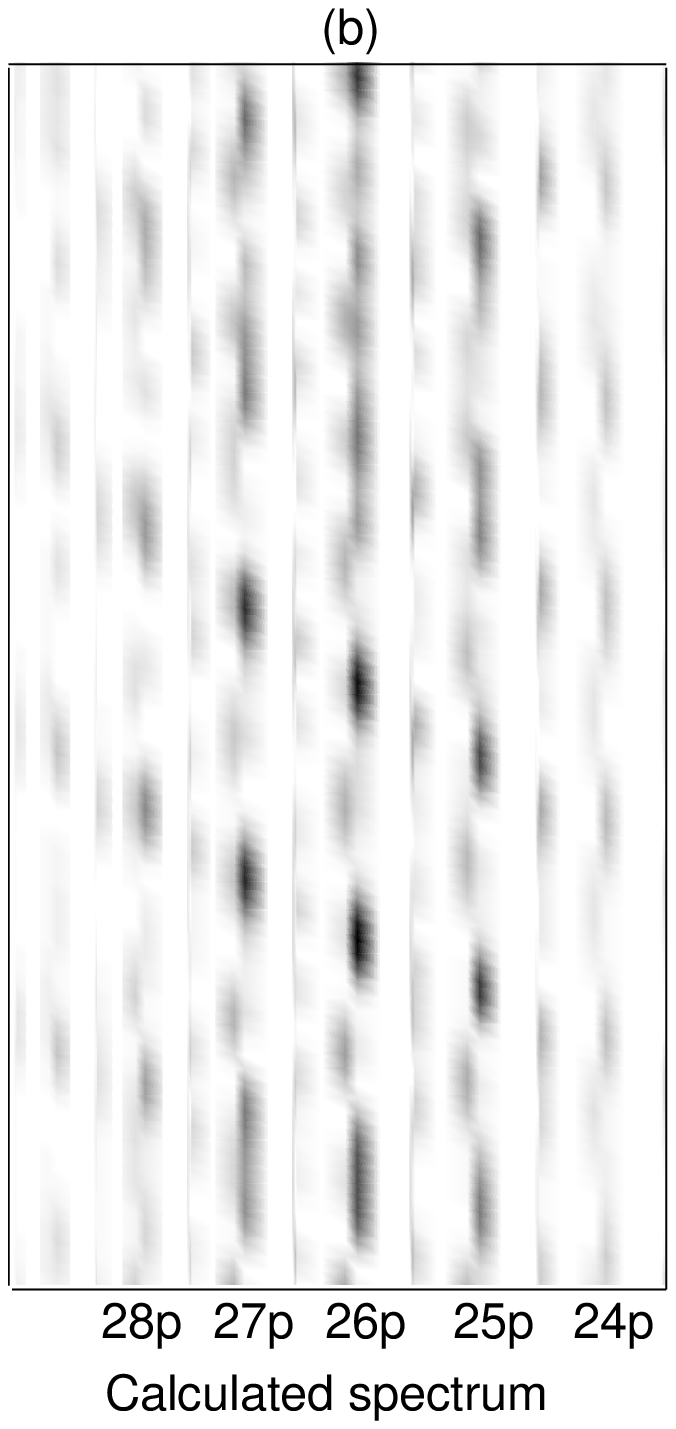}
\end{figure}

\end{multicols}

\end{document}